# The 7-channel FIR HCN Interferometer on J-TEXT Tokamak

**Wei Chen**[a,b], **L. Gao**[a,b*] **J. Chen**[a,b] **Q. Li**[a,b] **Z.J. Wang**[a,b] **G. Zhuang**[a,b]

[a] *State Key Laboratory of Advanced Electromagnetic Engineering and Technology, Huazhong University of Science and Technology, Wuhan, Hubei, China,430074*

[b] *Key Laboratory of Fusion and Advanced Electromagnetic Technology Ministry of Education, Huazhong University of Science and Technology, Wuhan, Hubei, China, 430074*

*E-mail*: gaoli@mail.hust.edu.cn

ABSTRACT: A seven-channel far-infrared hydrogen cyanide (HCN) laser interferometer has been established aiming to provide the line integrated plasma density for the J-TEXT experimental scenarios. A continuous wave glow discharge HCN laser designed with a cavity length 3.4 m is used as the laser source with a wavelength of 337 μm and an output power up to 100 mW. The system is configured as a Mach-Zehnder type interferometer. Phase modulation is achieved by a rotating grating, with a modulation frequency of 10 kHz which corresponds to the temporal resolution of 0.1 ms. The beat signal is detected by TGS detector. The phase shift induced by the plasma is derived by the comparator with a phase sensitivity of 0.06 fringe. The experimental results measured by the J-TEXT interferometer are presented in details. In addition, the inversed electron density profile done by a conventional approach is also given. The kinematic viscosity of dimethyl silicone and vibration control is key issues for the system performance. The laser power stability under different kinematic viscosity of silicone oil is presented. A visible improvement of measured result on vibration reduction is shown in the paper.

* Corresponding author.

## Contents

## 1. Introduction

A seven-channel far-infrared hydrogen cyanide (HCN) laser interferometer has been established aiming to provide the line integrated plasma density for the J-TEXT experimental scenarios. The system is configured as a Mach-Zehnder type interferometer [1]. The 7-channel FIR HCN interferometer contains the following parts: 1) continuous wave glow discharge HCN laser and collimation system He-Ne laser; 2) main parts of interferometer which contains diagnostic windows with 7-channel, tower as underprop and board as installed optical elements system, rotating grating, and so on; 3)detector and data acquisition and computer DAS system.

J-TEXT tokamak operates at electron density $10^{19}$~$10^{20}m^{-3}$, while cutoff density for HCN laser is $9.8\times10^{21}m^{-3}$. The laser is F-P waveguide laser [2][3] and operated in the $EH_{11}$ waveguide mode[4], while this is a low-loss linearly polarized mode which is achieved with the use of three tight fine metal wire. For our applications, the design power is about 100mW as inner diameter of waveguide tube $d = 50 \pm 2$ mm, cavity length is 3.4m, total loss a=3.5%. Control of the coupling loss $t_0$=6.5% is achieved with the use of 500 LPI Ni mesh. A precision thickness quartz crystal plate film whose reflection is less than 2% at 337μm is employed as output window. The glass tube is surrounded by dimethyl silicone, and its temperature is maintained at about 108℃.[5] It is important to note that too high kinematic viscosity will make the output is very unstable. The laser is assembled on a 5-meter optical table in the lower field area about 8-meter away from the device.

This paper will discribe interferometer system set-up first. In this part, optical layout of the interferometer is presented and data processing technique used in the system is discussed. The second part give the experiment results of interferometer including seven channel's line-integral density, profile of density by Abel inversion. The third part are key issues impacting the system performance, including power stability of laser source and vibration reduction. In the end of this paper, a brief conclusion is given.

## 2. Set up of system

### 2.1 Optical layout

The optical geometry of the system is shown in Figure 1. Components are mounted on four breadboards, three of which are rigidly supported by a single steel triangular frame while another breadboard is mounted on optical platform just before HCN laser's output window. Rotating grating with a modulation frequency of 10 kHz which corresponds to the temporal resolution of 0.1 ms, and collimation system is installed on the Board 4 shown in Figure 1. Table test operates on Board 4, which is made of steel. The other three breadboards, Board1~Board3 shown in Figure 1, are made of epoxy resin. The quartz beam-splitters (X-cut), BS2~BS8 shown in Figure 1, are reflecting in the visible to associated alignment. These beam-splitters are compensated to provide approximately uniform power to each detector. The optic viewing chords travel through the plasma vertically, their radial positions are r=21cm, 14cm, 7cm, 0cm, -7cm, -14cm, -21cm , respectively. Hollow dielectric waveguide is used in probe arm for beam transport, to reduce beam attenuation and control the beam divergence, as the vertical distance in Figure 1 is approximately 3m.

As shown in Figure 1, the vertical probe beam illuminates the plasma cross section from above. Below the tokamak, a beam-splitter combines it with the reference beam. The combined signal is fed into the detectors assembly, which consists of a DLATGS detector, a focusing mirror, and an adjustable mount. The DLATGS detectors exhibit low noise ($\sim 10^{-8} W/\sqrt{Hz}$ ) at room temperature, with a responsivity of ~21.5V/W.The diagnostic window (600-mm radial, 76-mm toroidal) consists of seven quartz plates (Φ=50mm) in perfect alignment, the plates have 70mm radial spacing corresponding to the spacing of detectors.

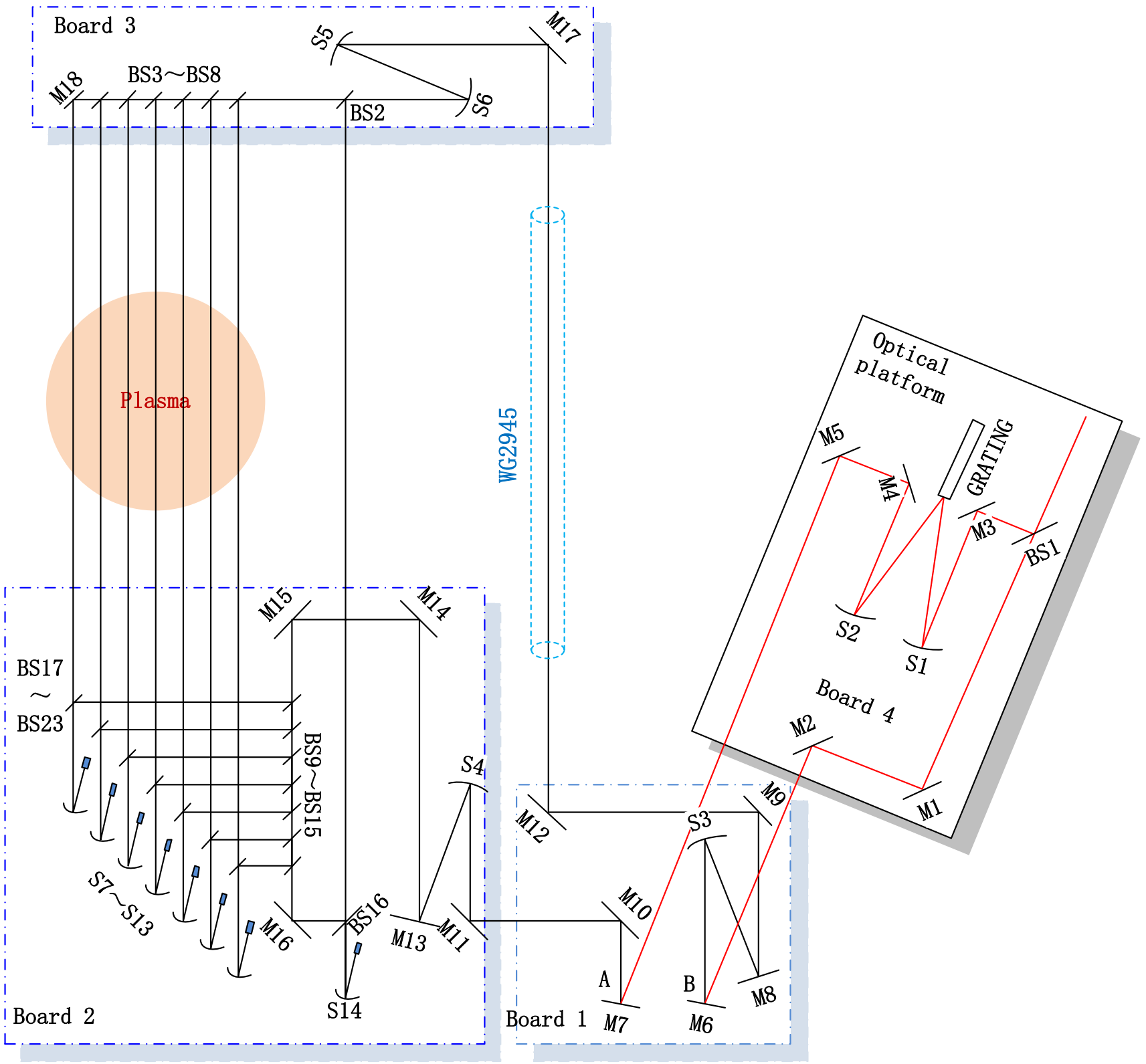

**Figure 1** Optical layout of the J-TEXT FIR interferometer.

All the parameters of the mirrors and focusing mirrors used in the interferometer can be calculated based on Gaussian theory.[6] To eliminate the vibration, the tower consists of three hollowed pillars are filled with more than 2 tons carborundum, and isolation technique with rubber bearing is also employed.

### 2.2 Phase comparator

There are two methods to obtain the phase shift: zero checked method and Fourier transform.[7] Flow chart of phase comparator based on FFT is shown in Figure 2. After checked, the digital signal passes through FIR filter to improve S/N by filtering some noises. FIR (Finite Impulse Response) filter is better than IIR (Infinite Impulse Response) filter. Because FIR filter is a linear phase filter, but IIR is in general not. Phase shift ranged in 0~π is obtained by conjugated multiplication of two IFFT signals. Jump point (from 0 jumping to π or vice versa) should be checked by setting a threshold to obtain a continuous density curve.

The phase shift is derived by the comparator with a phase sensitivity of ~0.06 fringe. A new phase comparator with phase-locked loop to lock the incoming signal [8] and a new counter to count phase jump based on FPGA technology is under tested to improve signal quality.

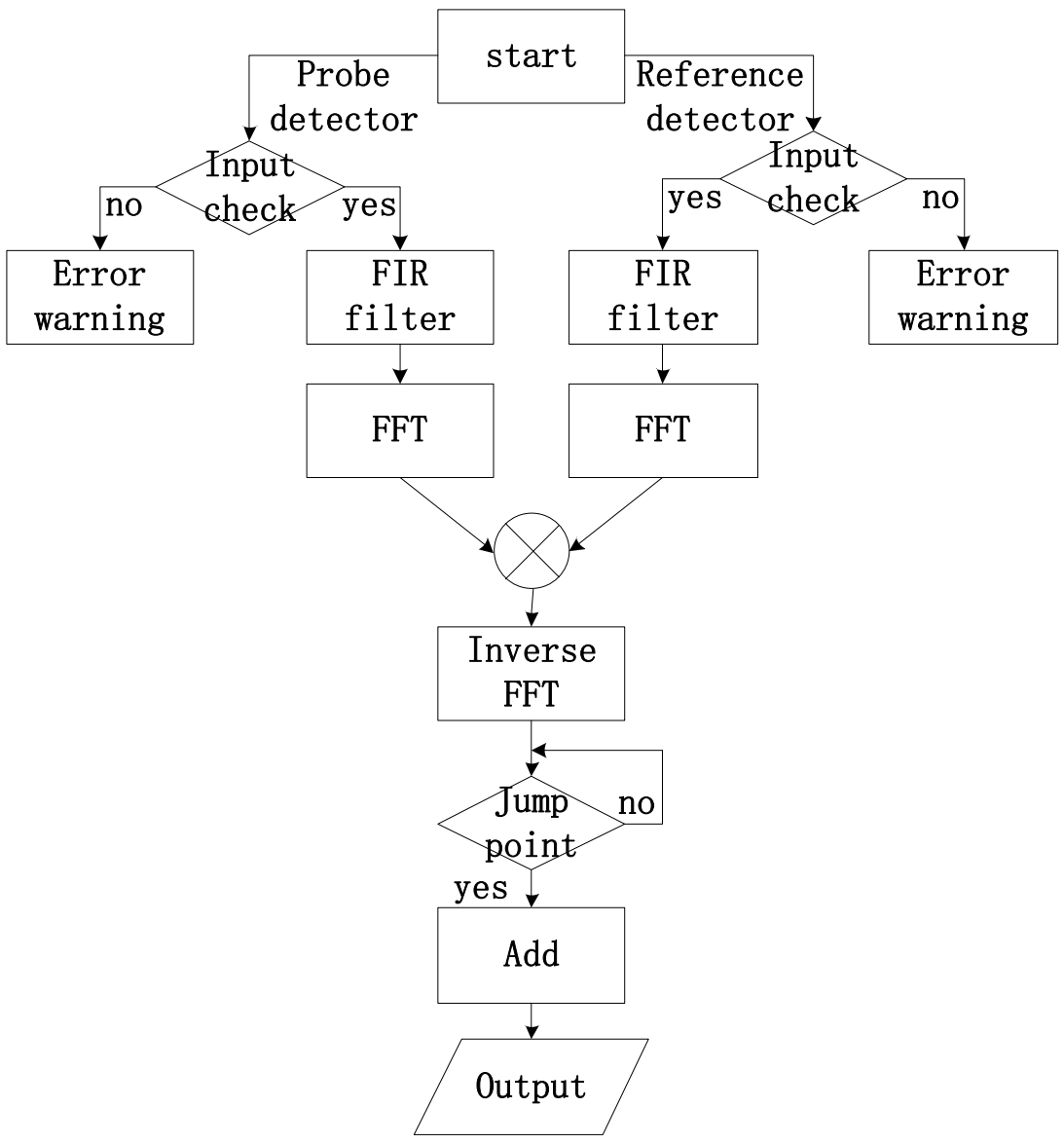

**Figure 2** Software phase comparator based on FFT.

## 3. Experiment result

The 7-channel vertical FIR HCN laser interferometer is constructed and it is a routine diagnostic on J-TEXT tokamak experiments since 2008. Typical time series data of J-TEXT tokamak ohmic discharge (waveforms of plasma current $I_p$, 7-channel line-integral density, vertical displacement DY, horizontal displacement DX, hydrogen-alpha $H_\alpha$, and gas puff) are shown in Figure 3(a), for an hydrogen plasma with $I_p$=200kA, $B_t$=1.67T and $\overline{n_e}(0) \approx 1.87 \times 10^{19} m^{-3}$. A polynomial fit procedure has been used, as shown in Figure 3(b). Data of 2.13s are used, shown as the solid square points. The line-integral density profile is approach to parabolic profile. The outward shift is consistent with magnetic measurements.

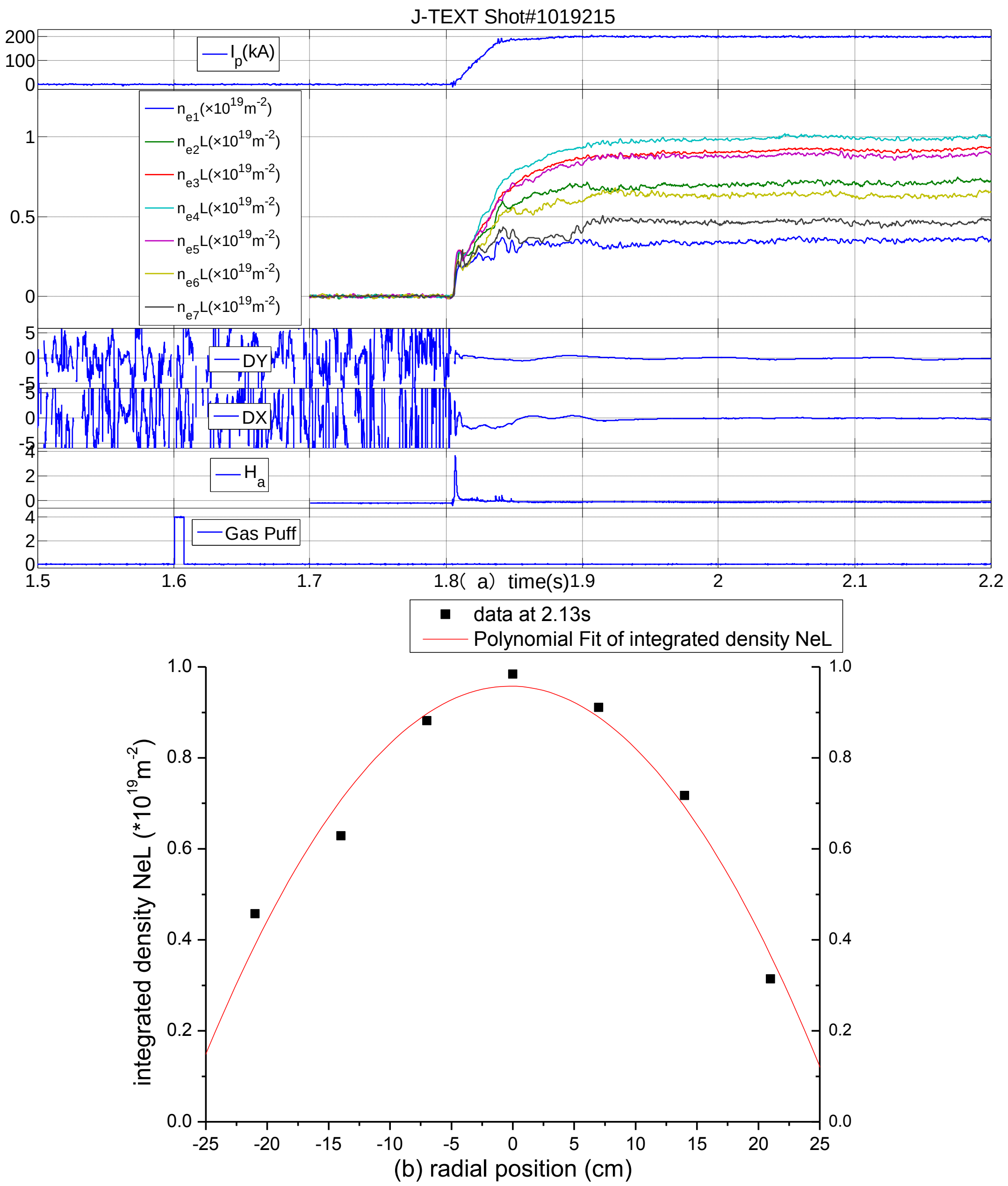

**Figure 3** a) Waveforms of plasma current $I_p$, 7-channel line-integral density $n_e L$, vertical displacement DY, horizontal displacement DX, hydrogen-alpha $H_\alpha$, and gas puff; b) The radial profile of line-integral density by polynomial fit.

No large deviation appears in any signal of the seven channel interferometer. In addition, the system has been phase calibrated. So the diagnostic results of FIR interferometer are relative credible when the plasma displacement is comparatively small. By performing a symmetrical Abel inversion, the local density profile is generated as shown in Figure 4. Here x-axis is radial position while y-axis is electron density with unit of $10^{19}m^{-3}$. The density is an approximately parabolic profile during the plateau region.

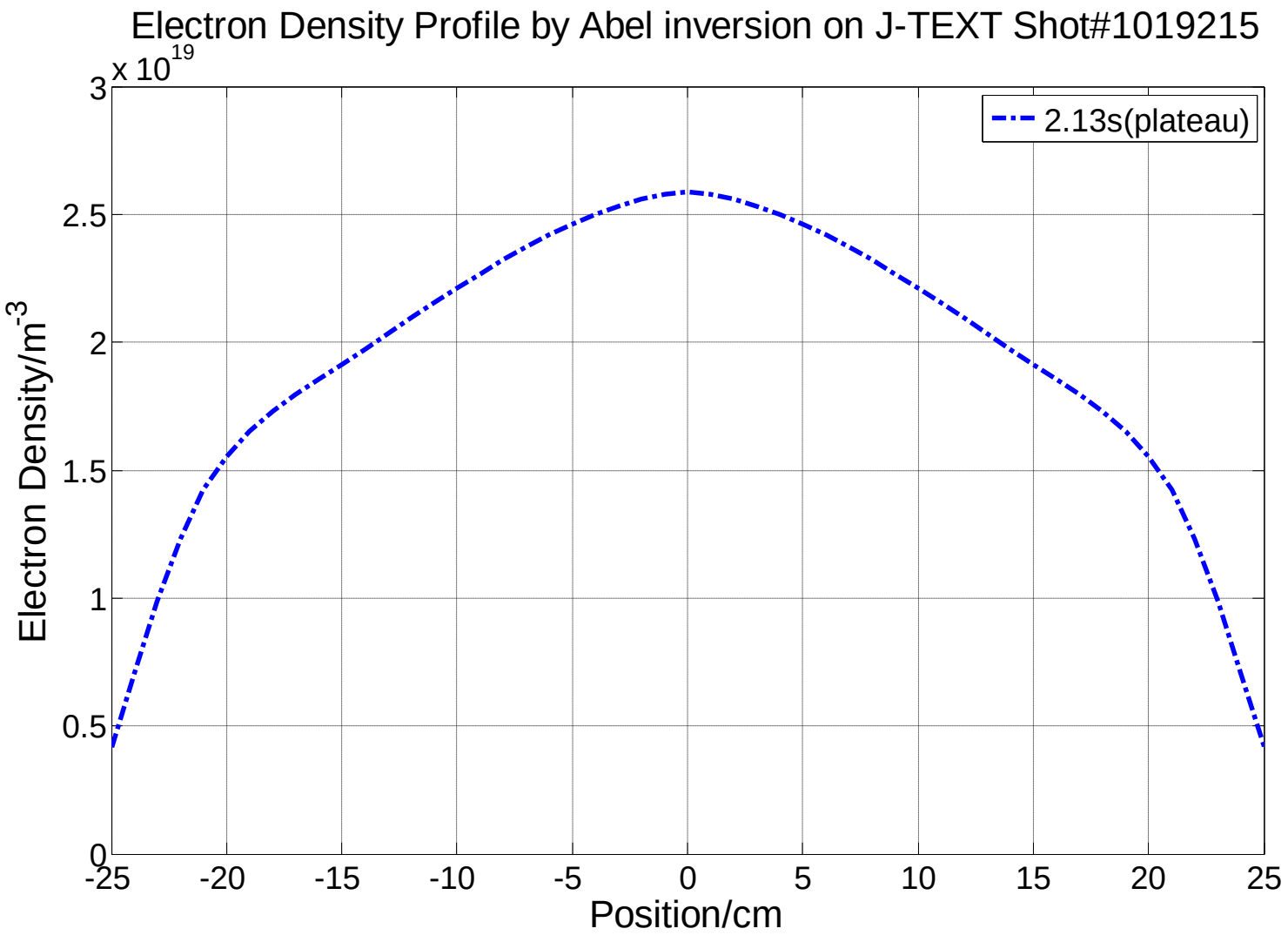

**Figure 4** Inverted electron density profile.

## 4. Key issues impacting the system performance

During development of the interferometer system, several factors were found impact the system performance, including output mode and the power stability of the laser, grating stability, and vibration of the system. Taking into account the specific circumstances of J-TEXT tokamak (the vibration of the system is very large in initial operation stage) and the limited space of the paper, only two more important issues, power stability of the laser and vibration, are discussed.

### 4.1 Laser source stability

The output power and power profile of HCN laser have been measured to optimize the laser output performance. Gaussian beam profiles are obtained and the output power is not very stable to 12.6% over a 1-h period without stabilization. As mentioned in reference [2][3][6], gas pressure and so on are important to laser performance, moreover kinematic viscosity of dimethyl silicone is key factor that can affect the performance of HCN laser greatly, which is not fully documented in the literature. If the kinematic viscosity of dimethyl silicone is too high, temperature distribution of the tube of wave glow is uneven in local, which will lead to power instability. Result of power stability is shown in Figure 5. Data is obtained after 3-hour's operation of the laser. After changing kinematic viscosity of silicone from 100 cSt to 50 or 20 cSt, the power stability of the laser's output is improved significantly. We define power stability time $t_{ps}$ as the time required for the power to decrease to 1/e (0.368) of its initial value. As shown in Figure 5, cavity length is adjusted when the power decrease to 1/e of its initial value. For 100cSt, $t_{ps}$ is about 180 minutes or 100minutes while $t_{ps}$ is up to 300 minutes for 50cSt and 20cSt.

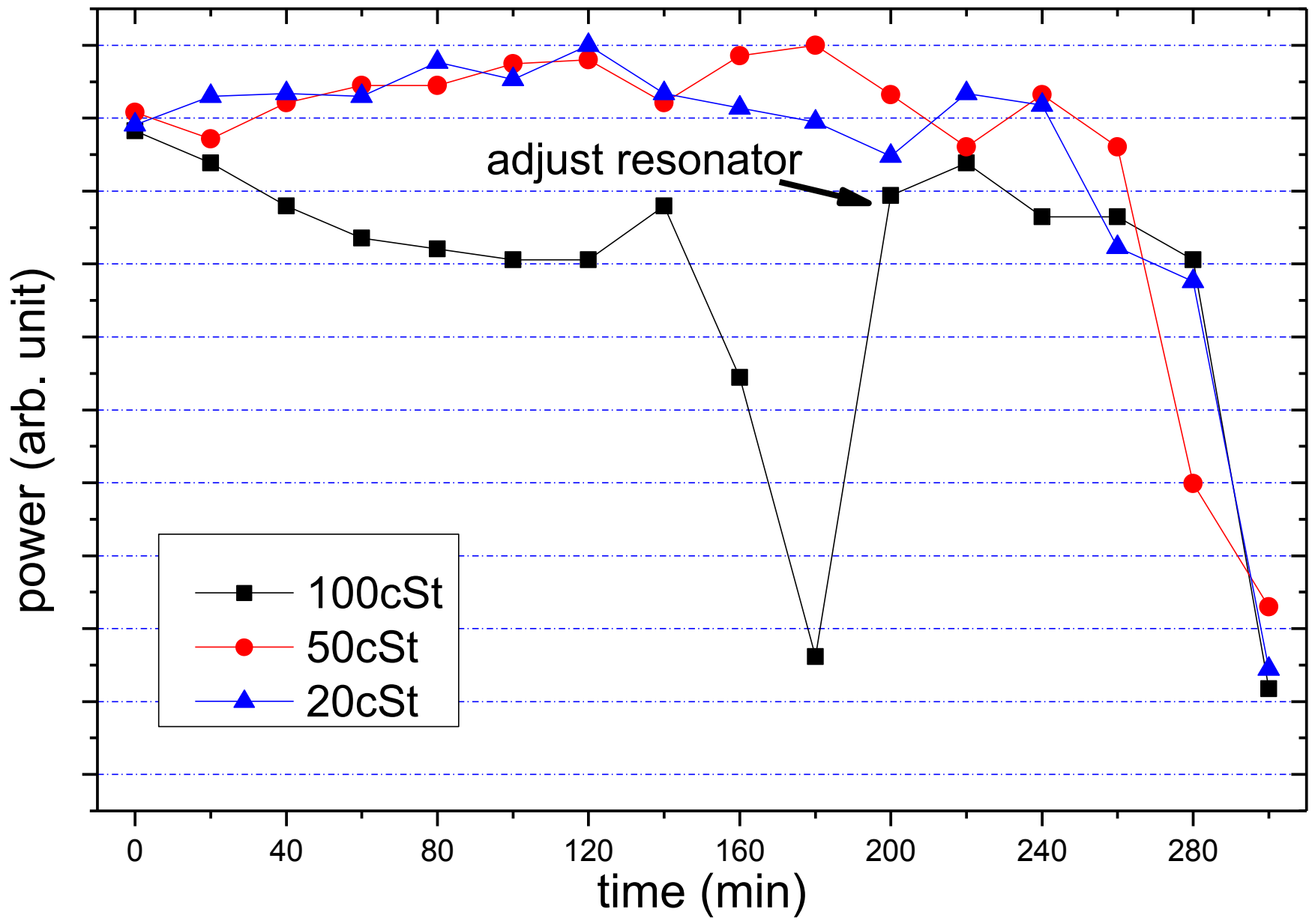

**Figure 5** The laser power stability under different kinematic viscosity of dimethyl silicone.

### 4.2 Reduction of the effect of vibration

At the early operation stage of the J-TEXT interferometer, the vibration affects the measurement greatly. To eliminate vibration of the system, the following measures are adopted: 1) the laser operates on marble optical platform instead of a normal optical table; 2) the HCN laser be moved far away from the device, at distances up to 8 meter; 3) let the reference beam pass through the Board2 in Figure 1; 4) reduce the length of strutting pieces of optical elements; 5) move the rotating grating to optical platform; 6) strengthen the support of breadboard. As shown in Figure 6, sensitivity to vibration can be reduced to a sufficiently low value after these changes.

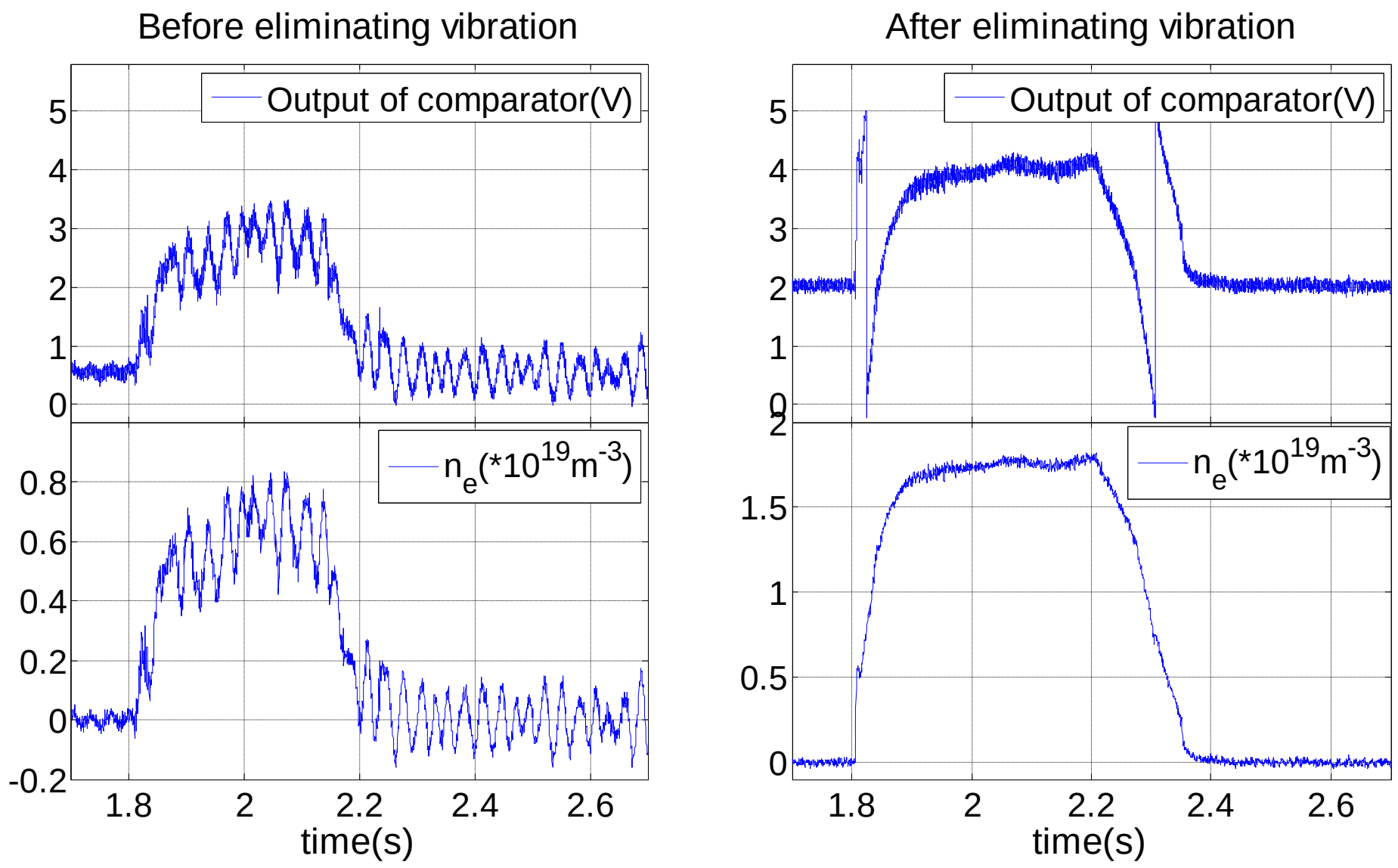

**Figure 6** Signal before and after reduction of the effect from vibration.

## 5. Conclusion

A 7 channel interferometer has been installed to obtain electron density profiles on J-TEXT tokamak. Felicitous kinematic viscosity of silicone is important to laser stability. Vibration of the system is reduced to a low level, which evidently improves the quality of phase shift signal. We will further improve the performance of the interferometer system and its S/N ratios.

## Acknowledgments

This research is supported by China National Key Basic Research Program (973 Project) No. 0215131007 and the National Natural Science Foundation of China (Grant No.11105056). Many thanks are due to Dr Ding Weixing from University of California, Los Angles, Dr Zhou Yan from Southwestern Institute of Physics for their help and suggestions.